\documentclass[%
 reprint,
superscriptaddress,
amsmath,amssymb,
aps,
]{revtex4-1}

\usepackage{graphicx}
\usepackage{dcolumn}
\usepackage{bm}
\usepackage{amsmath,amssymb}
\usepackage[usenames,dvipsnames]{color}

\begin{document}

\preprint{APS/123-QED}

\title{Persistent vs.\ arrested spreading of biofilms at
  solid-gas interfaces - the role of surface forces}

\author{Sarah Trinschek}
 \affiliation{Institut f\"ur Theoretische Physik, Westf\"alische Wilhelms-Universit\"at M\"unster\\ Wilhelm-Klemm-Strasse 9, 48149 M\"unster, Germany}
 \affiliation{Univ. Grenoble Alpes, LIPHY, F-38000 Grenoble, France}
  \affiliation{CNRS, LIPHY, F-38000 Grenoble, France}
 
\author{Karin John}%
\affiliation{Univ. Grenoble Alpes, LIPHY, F-38000 Grenoble, France}
\affiliation{CNRS, LIPHY, F-38000 Grenoble, France}

 \author{Sigol\`ene Lecuyer}%
\affiliation{Univ. Grenoble Alpes, LIPHY, F-38000 Grenoble, France}
\affiliation{CNRS, LIPHY, F-38000 Grenoble, France}
 
\author{Uwe Thiele}%
 \email{u.thiele@uni-muenster.de}
 \affiliation{Institut f\"ur Theoretische Physik, Westf\"alische Wilhelms-Universit\"at M\"unster\\ Wilhelm-Klemm-Strasse 9, 48149 M\"unster, Germany}
 \affiliation{Center of Nonlinear Science (CeNoS), Westf\"alische Wilhelms-Universit\"at M\"unster\\ Corrensstr. 2, 48149 M\"unster,  Germany}

\date{\today}

\begin{abstract}
We introduce and analyze a model for osmotically spreading biofilm colonies at solid-air interfaces that includes wetting phenomena, i.e.\ surface forces. The model combines a hydrodynamic description for biologically passive liquid suspensions with bioactive processes.
We show that wetting effects are responsible for a transition between persistent and arrested spreading and
provide experimental evidence for the existence of this transition for \textit{Bacillus subtilis} biofilms growing on agar substrates. 
In the case of arrested spreading, the biofilm is non-invasive albeit being biologically active. However, a small reduction in the surface tension of the biofilm is sufficient to induce spreading.


\end{abstract}

\pacs{Valid PACS appear here}

\maketitle

Biofilms are macro-colonies of bacteria enclosed in an extracellular matrix that form at diverse interfaces \cite{Donlan2002EID}. 
Cell proliferation and matrix  production by the bacteria result in lateral spreading
of the colony along the interface. Surprisingly, during the osmotic
spreading of biofilms on moist solid (agar)
substrates in contact with a gas phase, the spreading is not driven by
the active motility of individual bacteria but by growth processes and
the physico-chemical properties of the biofilm and the interfaces \cite{SAW+2012PNASUSA,ZSS+2014NJoP,DTH2014prsb}.
Within this mechanism, the biological production of polymeric matrix results in an osmotic 
flux of water from the agar into the biofilm that
subsequently swells and spreads out. As the spreading involves the
motion of a three-phase contact line between the viscous biofilm, the
agar and the gas phase, wetting phenomena \cite{DeGennes1985romp,BEI+2009RMP} are likely to
play an important role.
This idea is supported by experiments, that indicate a strong
dependence of biofilm spreading on their ability to
produce biosurfactants \cite{DRH+2006potnaos,LMB+2006aom,FPB+2012SM,KHC+2015fim,ARK+2009PNASU,
  BSZ+2009job,DFM+2015}.

To demonstrate the effect of surfactants, we perform osmotic
  biofilm spreading experiments using a \textit{Bacillus subtilis}
wild-type strain (WT) and a mutant strain with deficient
  production of surfactin ($\Delta$Srf) -- a natural
biosurfactant produced by WT \textit{B. subtilis}. Typically,
production is induced at high cell density by cell-to-cell
communication (quorum sensing) \cite{OSD+2014potnaos,GVK2015pb} and
plays a key role in surface motility \cite{SWK2006job, KL2003mm}.

\begin{figure}[b]
\includegraphics[width=0.4\textwidth]{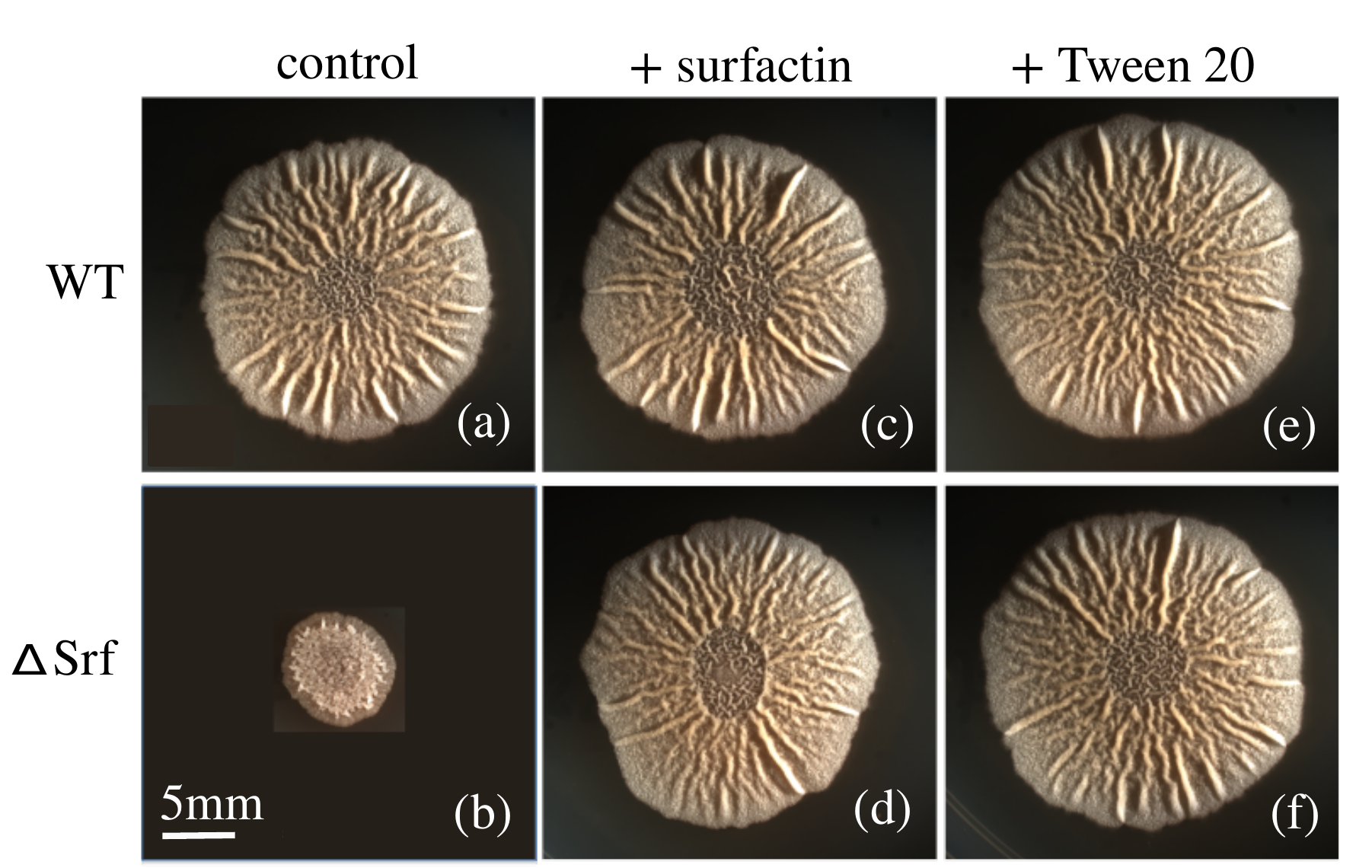}  
\caption{Experimental observation of the influence of surfactants on
  the transition between persistent and arrested biofilm spreading. (a) The \textit{B. subtilis} wild-type (WT) persistently spreads over the agar substrate, whereas (b) for the mutant strain with deficient surfactin production ($\Delta$Srf) spreading is arrested. An external addition of the surfactants
  (c,d) \textit{surfactin} or (e,f) \textit{Tween 20} enables the mutant
    strain to spread but does not affect the WT. }
\label{Fig_RescueExperiment}
\end{figure}
 
Bacterial suspensions and agar substrates are prepared as described in note \cite{experiment}. Agar plates with appropriate nutrient medium are inoculated with a small droplet of cell suspension, and biofilm growth and spreading is subsequently monitored. 
Figs.~\ref{Fig_RescueExperiment}(a) and~(b) show the colony of the
WT and the mutant strain, respectively, after 3 days of
incubation. The WT in (a) forms circular biofilms with a diameter
of about 2\,cm whereas the mutant strain without
\textit{surfactin} in (b) is not able to spread. The external addition of \textit{surfactin}
shortly after agar inoculation has no effect on the spreading of the
WT strain [Fig.~\ref{Fig_RescueExperiment}(c)], but restores a WT morphology in the
\textit{surfactin}-deficient strain [Fig.~\ref{Fig_RescueExperiment}(d)]. The WT phenotype can
also be recovered by adding the non-physiological
surfactants \textit{Tween 20} [Figs.~\ref{Fig_RescueExperiment}(e)-(f)] or \textit{Span80} (not shown),
which points at a physical role of \textit{surfactin} in the spreading
mechanism. 

To model the influence of wetting phenomena on biofilm
  spreading we supplement a hydrodynamic description of a thin film
of a biologically passive liquid suspension
\cite{TTL2013prl,Thiele2011ES,XTQ2015JPCM} by biomass growth
processes. The recently introduced model \cite{TJT2016ams} explicitly
includes surface forces, i.e., wettability, via a Derjaguin (or
disjoining) pressure and capillarity via the interface tensions.
Similar thin film models without wettability influences are
used to study early stage biofilm growth and quorum sensing
\cite{WK2012JEM}, osmotically driven spreading \cite{SAW+2012PNASUSA}
and the effect of surfactant production on the spreading of a
bacterial colony up a non-nutritive wall
\cite{ARK+2009PNASU}.

\begin{figure}[hbt]
\begin{center}
\includegraphics[width=0.47\textwidth]{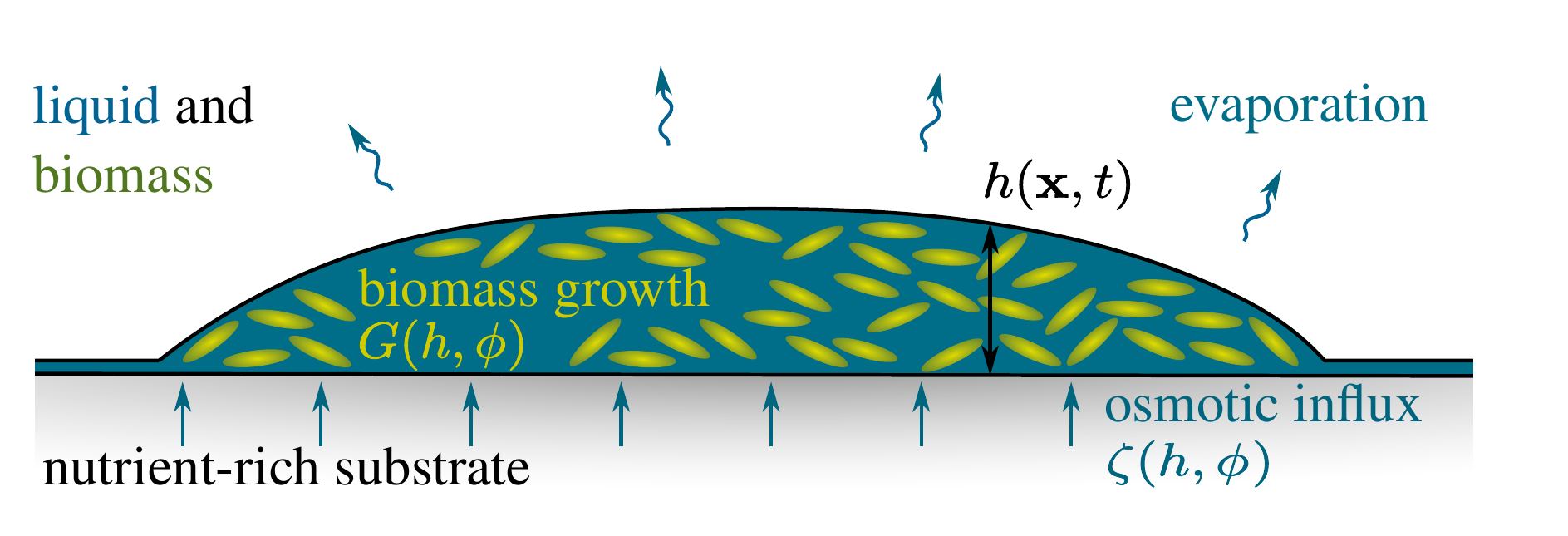}  
\caption{Sketch of the osmotically driven spreading  of a
    biofilm with the height profile $h(\mathbf{x},t)$. Osmotic pressure
    gradients are generated as bacteria consume water and nutrients to
    produce biomass via bacterial proliferation and matrix secretion,
    which is described by the growth term $G(h,\phi)$. This causes an
    osmotic influx of nutrient-rich water $\zeta(h,\phi)$ from the
    moist agar substrate into the biofilm.}
\label{Fig_1}
\end{center}
\end{figure}

Here, a biofilm of height $h(x,y,t)$ is modelled as
a mixture of solvent (nutrient-rich water) and of biomass (bacteria
and extracellular polymeric matrix) with the height-averaged
biomass concentration $\phi(x,y,t)$ (see Fig.~\ref{Fig_1}).  The free energy functional, that
determines all transport processes for the
passive suspension is
 \begin{align}\label{eqn:F}
  F[h, \phi] = \int \, [ \, &  f_w(h) + h f_m(\phi) + \tfrac{\gamma}{2} (\nabla h)^2 \,   ] \,   \mathrm{dA},
 \end{align}
 where $\gamma$ is the biofilm-air surface tension and
 $\nabla=(\partial_x,\partial_y)^T$ denotes the planar gradient
 operator. Furthermore, $\gamma_{SG}$ and $\gamma_{SL}$ are the
   solid-gas and solid-liquid interface energies.
A common choice for the wetting energy is \cite{Thiele2007,BEI+2009RMP}
\begin{equation} \label{eqn:f}
f_w(h) =   A (- \frac{1} {2h^2} + \frac{h_p^3}{5 h^5}  )  \, .
\end{equation}
which combines destabilising long-range van-der-Waals and stabilising
short-range interactions. Here, $h_p$ denotes the height of a thin wetting layer and 
\begin{equation} \label{eq:A}
A= \tfrac{10}{3}h_p^2 (\gamma - \gamma_{SG}+ \gamma_{SL})
\end{equation}
is the Hamaker constant, here expressed through the interface energies.
For a partially wetting biofilm-substrate-air combination,
minimizing Eq.\,(\ref{eqn:F}) gives the coexistence of a wetting
layer of height $h_p$ with steady droplets of equilibrium contact angle
$\cos\theta_\mathrm{eq} = 1+ f_w(h_p)/\gamma=  (\gamma_{SG} - \gamma_{SL})/\gamma$,
equivalent to the Young-Dupr\'e  equation \cite{DeGennes1985romp}.
The film bulk contribution
\begin{equation}
  f_m(\phi)= \frac{k_\text{B}T}{a^3}  [ \phi \ln (\phi) + (1-\phi) \ln (1-\phi)  ] \label{eq:g}
\end{equation}
 represents the entropic free energy of mixing of solute and solvent. We assume for simplicity, that biomass and solvent are represented by the same microscopic length $a$. $k_\text{B}T$ denotes the thermal energy.
 
The passive convective flux $ \mathbf{j}_\mathrm{conv}$ and diffusive flux $ \mathbf{j}_\mathrm{diff}$ 
are derived by applying a variational principle to the free energy
(\ref{eqn:F}) (for details see \cite{XTQ2015JPCM,TJT2016ams}):
\begin{align}
& \mathbf{j}_\mathrm{conv} =  \frac{h^3}{3 \eta} \nabla \left( \gamma \Delta h - \partial_h f_w \right) \\
& \mathbf{j}_\mathrm{diff} = -  D_\text{diff} h \phi  \nabla \left( \partial_{\phi} f_m \right)\, . 
\label{eqn:fluxes}
\end{align}
The composition-dependent viscosity $\eta$ of the biofilm \cite{HCS2004NRM, Sutherland2001m,LDB+2009bj} is given by $\eta = (1-\phi)\eta_0 + \phi\eta_b$, where $\eta_0$ and  $\eta_b$ denote the viscosity of solvent and biomass, respectively.
The biomass diffusivity is $D= \frac{a^2}{6 \pi \eta}$ consistent with 
the diffusion constant $D_\text{diff}= D \frac{k_\text{B} T}{a^3} = \frac{k_\text{B} T}{6 \pi a \eta}$.

The biomass multiplies by consuming nutrient-rich water following a bimolecular reaction $g\phi(1-\phi)$ with the growth rate constant g.
To account for processes such as nutrient and oxygen depletion \cite{ZSS+2014NJoP, DOP+2013Job}
we introduce a limiting amount of biomass $\phi_\mathrm{eq} h^\star$ by assuming a simple logistic growth law 
 \begin{equation}
 G(h, \phi) = g \phi (1-\phi) (1 - \tfrac{h \phi}{\phi_\mathrm{eq} h^\star})  \cdot  f_\text{mod}(h, \phi) \, .   
 \label{eqn:growth}
\end{equation}
$f_\text{mod}(h, \phi)$ 
\cite{fmod} 
modifies the growth law locally
for very small amounts of biomass. It ensures that at least one bacterial cell is needed for cell division and thus proliferation of biomass does not take place in the wetting layer.

 \begin{figure*}
\begin{center}
\includegraphics[width=\textwidth]{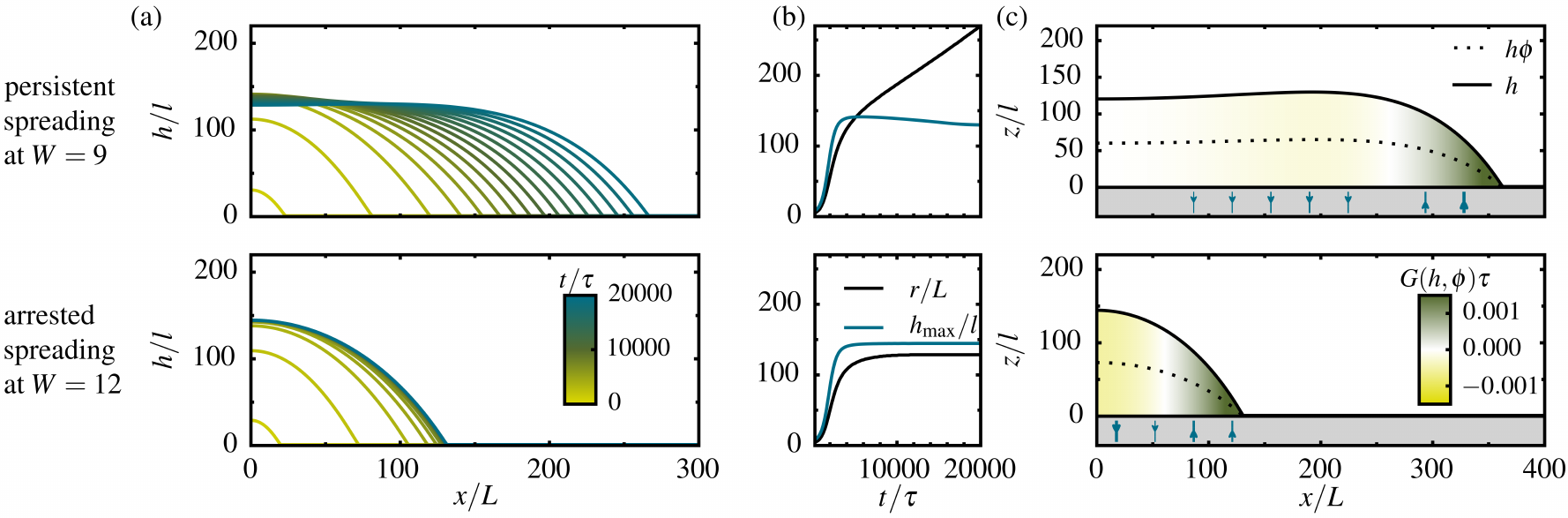}  
\caption{Comparison of persistently spreading biofilms (top row,
    at $W=9$) and arrested spreading of
  biofilms (bottom row, at $W=12$).  (a) Height profiles taken at
  equidistant times.  (b) Time evolution of the biofilm extension $r(t)$
  (solid black) measured at the inflection point of the height
  profile, and of the maximal film height $h_\mathrm{max}(t)$
  (solid blue).  (c) Bioactivity and osmotic influx for biofilms at
  a late time when all transients have decayed. The shading within the film indicates the
  bioactivity $G(h,\phi)$. The direction and strength of the effective
  osmotic flux $\zeta(h,\phi)$ is represented by the direction and
  thickness of the blue arrows below the biofilm.  Note that a time
  lapse of 1000\,$\tau$ corresponds to $\approx$5\,h. Remaining
  parameters are $\tilde g=0.01$ and $\tilde Q_\mathrm{osm}=100$.}
\label{Fig_SimResults}
\end{center}
\end{figure*}

Since the biomass cannot diffuse into the agar, biomass growth creates
an osmotic imbalance between the biofilm and the agar.
We assume, that the agar constitutes a large reservoir of nutrient rich
water at a constant osmotic pressure $\mu_{agar}$, corresponding to an
equilibrium water concentration $(1-\phi_{eq})$ in a flat biofilm.
The osmotic pressure in the biofilm, defined as the negative of the
variation of the free energy (\ref{eqn:F}) with respect to the
height $h$ at a fixed number of osmotically active particles $h\phi$,
is given by
\begin{equation}
  \mu_\mathrm{s} = - \frac{\delta F[h, \phi]}{\delta h} + \frac{\phi}{h} \frac{\delta F[h, \phi]}{\delta \phi} = - \partial_h f - g + \phi \partial_{\phi} g + \gamma\Delta h\, .
\label{eqn:piosm} 
\end{equation} 
The osmotic flux of water between agar and biofilm depends linearly on the osmotic pressure difference 
$\zeta(h, \phi)=  Q_\mathrm{osm} \left(    \mu_\mathrm{s} - \mu_\mathrm{agar}  \right )$,
with $Q_\text{osm}$ being a positive mobility constant.

Biomass growth and osmotic flux are incorporated into the model as two non-conserved terms $G(h, \phi)$ and $\zeta(h, \phi) $ which results in the following evolution equations for the effective layer thicknesses of liquid $h$ and biomass $h\phi$
\begin{align} 
&\partial_t h = - \nabla \cdot \mathbf{j}_\mathrm{conv} + \zeta(h,\phi) \\
&\partial_t(h\phi) = - \nabla \cdot(\phi \mathbf{j}_\mathrm{conv}
+\mathbf{j}_\mathrm{diff}   )+ h G(h,\phi) \, . \label{eqn:biofilmeqn}
\end{align} 
Note that the conserved part of the dynamics can also be given in gradient dynamics form \cite{TJT2016ams, XTQ2015JPCM, WTG+2015apa}.

To facilitate the model analysis, 
we introduce vertical and horizontal length scales $l=h_p$ and
$L=(\gamma/\kappa)^{1/2} l$, respectively, with $l \ll L$, the time
scale $\tau= L^2\eta_0/\kappa l$ and the energy scale
$\kappa=k_\text{B} Tl/a^3$. This gives dimensionless growth
  rate $\tilde g = g \tau$, osmotic mobility $\tilde Q_\mathrm{osm} =
  Q_\mathrm{osm} \tau \kappa/l^2$ and wettability parameter
\begin{equation}
W=  \frac{A}{\kappa l^2} = \frac{A}{k_\text{B} T} \frac{a^3}{l^3}
\label{eq:W}
\end{equation}
that measures the relative strength of the wetting energy \cite{PT2006pof1} as compared to the entropic free energy of mixing. Larger values of $W$ correspond to a less wettable substrate and result in larger equilibrium contact angles.

Throughout the analysis we fix the maximal amount of biomass that can
be sustained by the substrate to $h^\star \phi_\mathrm{eq}=60$, the
equilibrium water concentration to $(1-\phi_{\mathrm{eq}})=0.5$ and
the ratio of the viscosities of biomass and fluid to
$\frac{\eta_b}{\eta_0}=20$. The biofilm spreading behavior is
  studied depending on the growth rate $\tilde g$, the wettability
  parameter $W$ and the osmotic mobility $\tilde Q_\text{osm}$.
Comparing with the typical biofilm height of $30\,\mu$m measured in
\cite{SAW+2012PNASUSA} and using viscosity and surface tension of
water ($\eta_0=10^{-3}$\,Pa s, $\gamma=70$\,mN/m) as well as the typical
solvent/biomass length scale $a=100$\,nm, this results in the vertical
length scale $l=h_p=250$\,nm, the lateral length scale $L=70\,\mu$m
and the time scale $\tau=20$\,s. With the above scales, a wettability
parameter $W=10$ corresponds to an equilibrium contact angle of
$0.5\,^\circ$, comparable to the dynamic contact angle measured in
\cite{SAW+2012PNASUSA}.

We analyze Eqs.\,(\ref{eqn:F}-\ref{eqn:biofilmeqn}) for a
two-dimensional geometry (biofilm ridges instead of circular colonies)
with no-flux boundaries employing numerical time simulations (finite element modular toolbox DUNE-PDELAB \cite{BBD+2008C1, BBD+2008C2} as previously used in \cite{TJT2016ams,WTG+2015apa}) and continuation techniques \cite{DWC+2014ccp} (software package \textit{Auto-07p} \cite{DOC+2007}).

Our model (\ref{eqn:F}-\ref{eqn:biofilmeqn}) reproduces the
  non-equilibrium transition between persistently spreading biofilms and
  arrested spreading: On the one hand, at relatively high wettability
  (lower $W$, Fig.~\ref{Fig_SimResults} (top, a) and (top, b)) the
  biofilm initially rapidly swells vertically and horizontally until a
  stationary film height is reached. Subsequently, it only spreads horizontally
  with a constant speed and shape of the biofilm edge. This
  qualitatively reproduces common experimentally observed behaviour
  \cite{SAW+2012PNASUSA,DTH2014prsb}.
Fig.~\ref{Fig_SimResults}~(top, c) shows a snapshot of a spreading biofilm
at a late time when all transients have decayed.  Far from the advancing edges, the biofilm has reached the limiting amount
of biomass and the biomass concentration corresponds to the
equilibrium value $\phi_\mathrm{eq}$ so that all bioactive processes
are in a dynamical equilibrium. At the edges, biomass growth takes
place and causes an osmotic imbalance that results in a strong influx
of water into the biofilm.

On the other hand, at lower wettability [larger $W$,
  Fig.\,\ref{Fig_SimResults} (bottom, a and b)], biofilm spreading is
  arrested. Again, the biofilm initially  rapidly swells, however, in
  contrast to the case of higher wettability, it soon evolves towards a steady profile of fixed extension and contact angle. Note that the steady biofilm drops are still bioactive - Fig.\,\ref{Fig_SimResults} (bottom, c) shows that biomass is being produced at the biofilm edges where $G>0$ and is degraded at
the centre where $G<0$ as there the biomass exceeds the limiting amount
$\phi_\mathrm{eq} h^\star$. This is possible as hydrodynamic and
diffusive fluxes within the biofilm and osmotic fluxes between agar and biofilm
rearrange biomass and water such that their profiles are stationary.
\begin{figure}
\begin{center}
\includegraphics[width=0.5\textwidth]{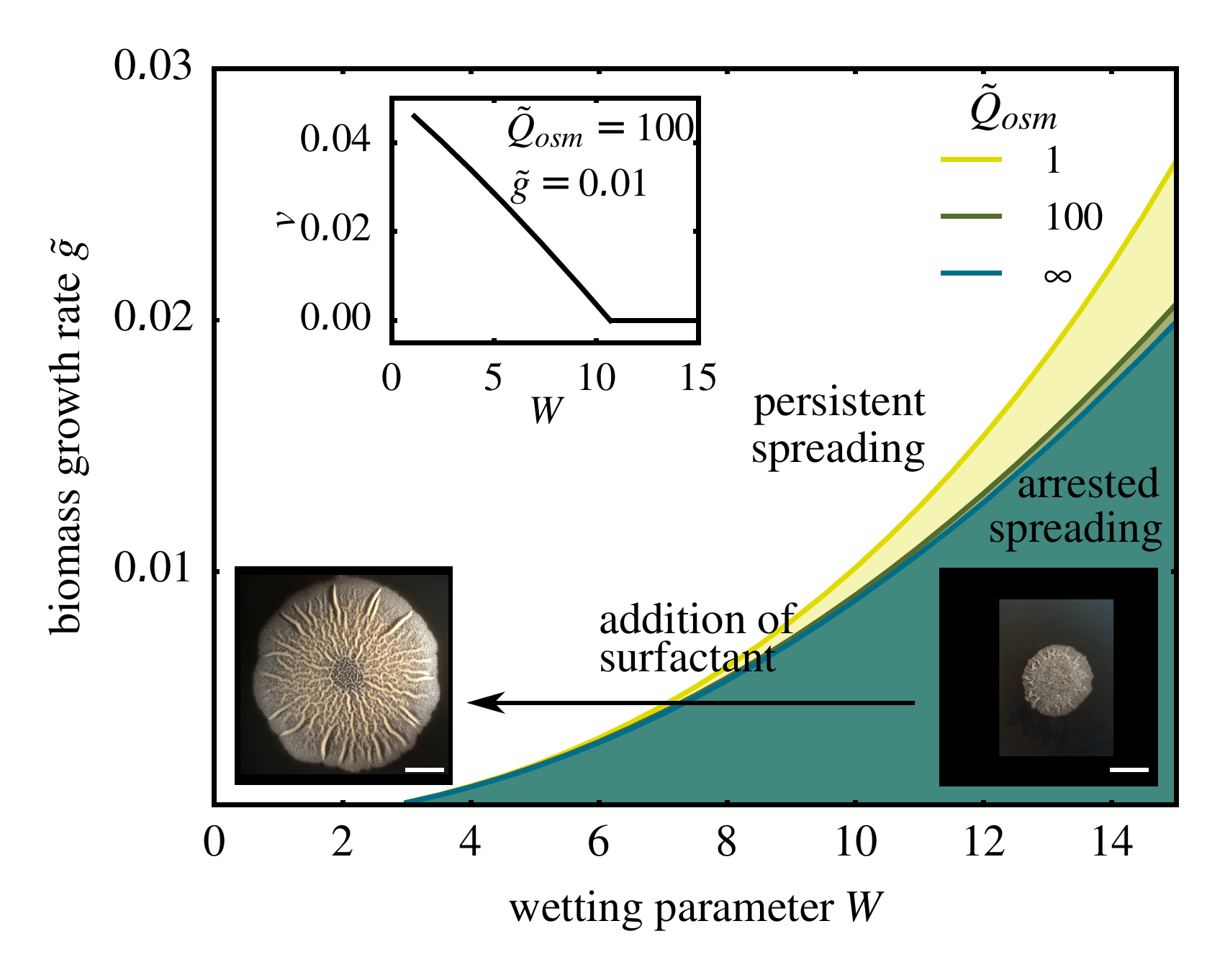} 
\caption{Spreading behavior of the biofilm in the $\tilde g$-$W$
  parameter plane for various values of the osmotic mobility
  $\tilde Q_{osm}$ as indicated in the legend. In the shaded regions
  biofilm spreading is arrested, i.e., it reaches a steady
  profile while outside spreading is persistent. The inset gives the
dependence of spreading speed $v$ on
  wettability $W$ for $\tilde g=0.01$ and $\tilde Q_{osm}=100$. A
  speed of $v=0.01$ corresponds to an actual spreading speed of
  0.1\,mm/h, comparable to the experiment in
  \cite{SAW+2012PNASUSA}. (scale bar: 5\,mm)}
\label{Fig_phasediagram}
\end{center}
\end{figure}

The spreading behavior in dependence of wettability parameter
  $W$ and the biomass growth rate $\tilde g$ is summarized in the
  non-equilibrium phase diagram presented in
  Fig.~\ref{Fig_phasediagram}.  At constant $\tilde g$, corresponding,
  e.g., to a specific bacterial strain, spreading of the biofilm is
  arrested at low wettability (high value of $W$).  However, as adding
  a surfactant lowers the biofilm surface tension and, consequently,
  the parameter $W\sim\gamma$ [cf.~Eqs.\,(\ref{eq:A}) and
  (\ref{eq:W})], it can trigger a transition from a non-invasive
  biofilm with arrested spreading to a persistently spreading biofilm
  -- in agreement with the experimental results
  in Fig.~\ref{Fig_RescueExperiment} and Ref.~\cite{LMB+2006aom}. 

This transition only slightly depends on the osmotic mobility
$\tilde Q_\mathrm{osm}$ (see Fig.~\ref{Fig_phasediagram}) and
  one may consider the limiting case of an instantaneous osmotic
  solvent transfer between agar and biofilm
  (i.e.,~$\tilde Q_\mathrm{osm} \gg 1$). There, the model reduces to a
  one-variable model for the evolution of the biofilm height 
  \cite{redModel}
  , and still reproduces all relevant experimental
  features.  Even for this infinitely fast osmosis, the parameter region of
  arrested spreading is only slightly smaller than for finite
  $\tilde Q_\mathrm{osm}$. This indicates that thermodynamic forces
  (surface forces, entropic forces) rather than time scales of
  transport and bioactive processes are dominant in the determination
of the transition between steady and invasive biofilms.
 
\label{sec:conc}
To summarize, we have presented a simple model for the osmotic
spreading of biofilms that grow at solid-air interfaces. The model
adds bioactive processes into a hydrodynamic approach and explicitly includes wetting effects.
In consequence, it has allowed us to study the interplay between biological growth processes and passive surface forces. Our results have confirmed within a thermodynamically consistent
framework that wetting crucially effects the spreading dynamics
(invasiveness) of biofilms and has therefore provided a qualitative
understanding of the experimentally observed transition between
arrested and persistent spreading that occurs upon addition of external surfactants in a surfactin-deficient \textit{B. subtilis} strain. 

Our framework is limited to the biologically early stages of biofilm growth since it
neglects all vertical gradients in the biofilm. However, it is well suited
to model the dynamics of the spreading biofilm edge. In future
  extensions, one may incorporate the auto-production and dynamics
of surfactants in the biofilm to consistently study the influence of
Marangoni flows on the spreading dynamics \cite{ARK+2009PNASU, DRH+2006potnaos, FPB+2012SM}.
As such flows are known to cause fingering
instabilities in spreading surfactant-covered droplets \cite{MT1999pof} these
mechanisms should be explored in connection to the branched structure
described for some biofilm colonies.
 

\acknowledgments We thank the lab of R.~Losick at Harvard University
for bacterial strains, the DAAD and
Campus France (PHC PROCOPE grant~35488SJ) for financial
support. LIPhy is part of LabEx~Tec~21 (Invest.\ l'Avenir,
grant~ANR-11-LABX-0030).

\bibliography{./LiteraturBiofilms.bib}

\begin{thebibliography}{37}
\bibitem [{fmo()}]{fmod}%
  \BibitemOpen
  \href@noop {} {}\bibinfo {note} {The explicit form of $f_\text{mod}(h, \phi)$
  is chosen as \begin{equation} f_\text{mod}(h, \phi) = (1 -
  \frac{\psi_{2,u}}{\phi_2 h}) (1- \exp(\phi_\mathrm{eq}h_p - \phi h))
  \end{equation} but other choices for $ f_\text{mod}(h, \phi)$ with the same
  fix point structure do not change the results qualitatively.}\BibitemShut
  {Stop}%
\bibitem [{red()}]{redModel}%
  \BibitemOpen
  \href@noop {} {}\bibinfo {note} {If the influx of water is fast as compared
  to biomass growth, but not as fast as the hydrodynamic fluxes, one can reduce
  the model to an effective description of the biofilm height as the large
  biofilm droplets are always in 'osmotic equilibrium' $\mu_\mathrm{s} =
  \mu_\mathrm{agar} $ with the substrate on the time scale of the biomass
  growth with \begin{equation} \zeta(h, \phi) = \frac{1}{1-\phi_{eq}} G(h,\phi)
  \,. \notag \end{equation} The evolution equation for the total height reduces
  to \begin{equation} \partial_t h = \nabla \cdot \left[ \frac{h^3}{3 \hat
  \eta} \nabla (\partial_h f - \Delta h) \right] + \frac{1}{1-\phi_{eq}} G(h,
  1-\phi_{eq}) \notag \end{equation} with a constant viscosity $\hat \eta =
  \phi_{eq} + k_\eta(1-\phi_{eq})$.}\BibitemShut {Stop}%
\end{thebibliography}

\end{document}